\documentstyle[12pt]{article}
\include{epsf}
\def\lo{\langle 0 |}
\def\ro{ | 0 \rangle }
\def\rpi{|\pi(q) \rangle }

 \def\rro{| \rho(q) \rangle }
 \def\fro{ f_{\rho}}
 \def\mro{m_{\rho}}
 \def\emu{\varepsilon_{\mu}^{(\lambda)} }
 
\def\ealpha{\varepsilon_{\alpha}^{(\lambda)} }
 \def\ebeta{\varepsilon_{\beta}^{(\lambda)} }
\def\gmmu{\gamma _{\mu}}
\def\gmnu{\gamma_{\nu}}

\def\gmf{\gamma _{5}}

\def\la{\langle }
\def\ra{ \rangle }
\def\nableft{ \stackrel{\leftarrow}{\nabla}}
\def\nabright{ \stackrel{\rightarrow}{\nabla}}
\def\nableftright{\stackrel{\leftrightarrow}{\nabla}}
\def\wf{\Psi(\xi,k_{\perp}^{2}) }
\def\wfu{\Psi(u,k_{\perp}^{2}) }
 \def\k{\vec{k}_{\perp}^2}
 
\newcommand{\beq}{\begin{equation}}
\newcommand{\eeq}{\end{equation}}
\newcommand{\bea}{\begin{eqnarray}}
\newcommand{\eea}{\end{eqnarray}}

\begin{document}
\renewcommand{\thefootnote}{\fnsymbol{footnote}}
                                        \begin{titlepage}
\begin{flushright}
hep-ph/9612425
\end{flushright}
\vskip1.8cm
\begin{center}
{\LARGE
Hard diffractive electroproduction,  \\
 \vskip0.7cm
transverse momentum distribution \\
 \vskip1.1cm
and QCD vacuum structure}

\vskip1.5cm
 {\Large Igor~Halperin} 
and 
{\Large Ariel~Zhitnitsky}
\vskip0.2cm
        Physics and Astronomy Department \\
        University of British Columbia \\
        6224 Agriculture Road, Vancouver, BC V6T 1Z1, Canada \\  
     {\small e-mail addresses: 
higor@axion.physics.ubc.ca \\
arz@physics.ubc.ca }\\
\vskip1.5cm
{\Large Abstract:\\}
\end{center}
\parbox[t]{\textwidth}{
We study the impact of the "intrinsic" hadron transverse momentum on the pre-asymptotic behavior of the diffractive electroproduction of longitudinally polarized $ \rho$-meson. Surprisingly, we find the onset of the asymptotic regime in this problem to be rather low,
 $Q^2 \simeq 10 \; GeV^2 $ where power corrections due to the transverse momentum do not exceed 20 \% in the amplitude. This drastically contrasts with exclusive amplitudes where the asymptotics starts at much higher $ Q^2 = 50 - 100 \; GeV^2 $. The sources of such unexpected behavior are traced back to some general (the quark-hadron duality) as well as more silent (properties of higher dimensional vacuum condensates) features of QCD.

 }

\vspace{1.0cm}
\begin{center}
{\em submitted to Phys. Rev. D }
\end{center}
                                                \end{titlepage}

\section{Introduction}

Over the last couple of years, following the paper \cite{BFGMS}, the diffractive 
electroproduction of the light 
vector mesons has been investigated within the framework
of perturbative QCD. Since this paper a few more important results have been
obtained. First of all it is a rigorous 
QCD-based proof \cite{Strikman} that an amplitude of 
diffractive electroproduction of vector mesons has required factorization
properties.  The second important step in the same direction
 is an introduction of a new type of   functions (the so-called antisymmetric
structure functions) \cite{Radasym} which are relevant objects for   such kind of problems.
Therefore, the standard  perturbative QCD (pQCD) approach
to this   new class of problems which are currently intensively studied experimentally \cite{DESY} has 
obtained a solid basis for the future investigation. 
\par 
 We should note that the diffractive 
electroproduction had been discussed many times before the
 recent paper \cite{BFGMS},
see e.g. \cite{xxx} and references therein. However, all previous approaches to the problem
were mainly    based on some kind of   quark models
with inevitable to such methods
 specific assumptions  not following
  from QCD. This is certainly a self-consistent approach for a
heavy quark system (where the size
of a hadron is of the order of $1/m_Q$ and is always small by kinematical reasons),
 but not for the light hadrons, where 
the problem of the separation of  small and large distance physics
cannot even be formulated in an appropriate way within the quark model. 
 Let us remind that
the main idea of the QCD-based approach  
 is just such a separation of the large and small
distance physics, the so-called
Wilson Operator Product Expansion (OPE).
 At small distances one can use the standard
perturbative expansion due to the asymptotic freedom and
smallness of the coupling constant. All nontrivial,
large distance physics is hidden in 
nonperturbative  hadron wave functions (WF's) in this approach.
They cannot be found
by the perturbative technique, but rather  should be extracted
from  elsewhere. Therefore, the problem is reduced to the analysis
of the bound states within OPE.

The problem of bound states in the
relativistic quantum field theory  with
large coupling constant is, in general, an extremely difficult
problem. Understanding the structure of the
bound state is a very ambitious goal which
assumes the 
solution of a whole spectrum of tightly connected problems, such as 
confinement,
chiral symmetry breaking phenomenon,   and many others which
are greatly important in the low energy region. Fortunately, a great deal of information can be obtained even in absence of such a detailed knowledge. This happens in high energy processes where the only needed nonperturbative input are the so-called light-cone WF's with a minimal number of constituents. In this case a problem becomes tractable within existing nonperturbative methods.
  
 We have therefore a well-formulated problem  of
the diffractive electroproduction of vector mesons. 
The formulation is based on the solid background of QCD.
However, calculations of the pQCD approach refer, strictly speaking, only to asymptotically high  energies \cite{BFGMS}.
Therefore, as usual,
the {\bf main question } remains: at what energies do the asymptotic formulae of
Ref.\cite{BFGMS} start to work? The prime goal
of the present work is to attempt to answer this question.
We should note that there are two very different types of pre-asymptotic corrections 
to the asymptotically leading formula \cite{BFGMS}.
First, there exist the so-called shadowing (rescattering) corrections
\cite{Maor} which we do not discuss in this paper at all. Our concern is rather with 
a second kind of corrections which
are related to the internal hadron structure of the vector meson.
This kind of corrections was discussed for the first time
in Ref.\cite{FKS}, and we will discuss the relation
of the paper \cite{FKS} to our results  later on.

However, before deepening in details, we would like to recall
that a very similar question about an applicability of the
QCD-based approach
to exclusive amplitudes
has been discussed   during the last fifteen years . The history of this development is very instructive
and we believe it is worthwhile to mention it  shortly here.

As is known, at asymptotically high energies
  the  parametrically leading contributions
to hard exclusive processes can be expressed in terms of the so-called
distribution amplitude   $\phi(x)$ \cite{Brod},
which itself can be expressed as an integral
$\int d^2 \vec{k}_{\perp}\psi (\k,x)$ with nonperturbative
wave function   $\psi(\k , x)$,
see reviews \cite{Brod1},\cite{CZ} 
  for details. 
 Distribution amplitudes  $\phi(x) $ depend only on longitudinal variables
$x_i$ and not on transverse $\k$ ones. The same is true
for inclusive reactions where structure functions depend on $x$,
but not on $\k$\footnote{ The formal reason for
that can be seen from the following arguments. At large energies the quark
and antiquark are produced at small distances 
$z\sim 1/Q\rightarrow 0$, where $Q$ is typical
large
momentum transfer. Thus one can neglect the $z^2$
dependence everywhere and one should concentrate
on the one variable $zQ\sim 1$ which is of the order of one.
One can convince oneself that the standard Bjorken
variable $x$ is nothing but Fourier conjugated to $zQ$.}.
 Thus, any dependence on $\k$ gives
some power corrections to the leading terms.
Naively one may  expect that these corrections should be small
enough already in the   few $GeV^2$ region.
These expectations are mainly based 
not on a theoretical analysis but rather on the phenomenological observation
that  the dimensional
counting rules,  proposed in early seventies (see  \cite{Matveev} ),
agree well with the experimental data such as the
 pion and nucleon form factors,
large angle elastic scattering
cross sections and so on. This agreement 
can be interpreted as a strong argument that power corrections are small in the   few $GeV^2$ region.
 
 However, in mid eighties the applicability of the approach
\cite{Brod}  at experimentally
accessible momentum transfers was questioned \cite{Ditt},\cite{Isgur}.
In these papers
 it was demonstrated, that the perturbative, asymptotically 
leading contribution is much smaller than the nonleading "soft"
one.
Similar conclusion, supporting this result, came from
the different side, from  the QCD sum rules
\cite{Rad},\cite{Smilga}, where the direct
calculation of the form factor has been presented
 at $Q^2\leq 3GeV^2$.
  This method
 has been extended later for 
  larger $  Q^2\leq 10 GeV^2$ \cite{Rad1},\cite{BH}
with the same qualitative result:
the soft contribution is more important in this intermediate region
than the leading one.

Therefore, nowadays it is  commonly accepted    that 
the asymptotically leading contribution to the exclusive amplitudes
cannot provide
the experimentally observable absolute
values at accessible momentum transfers. 
If we go along with this proposition, then the natural question   arises: 
 How   can one
explain the very good agreement between the
experimental data and  dimensional counting rules
if the asymptotically leading contribution cannot explain the data
for experimentally accessible energies?
A possible answer was suggested recently \cite{ChibZh} 
and can be formulated in the following way: very unusual properties
 of the transverse momentum distribution
of a hadron lead to the
{\bf mimicry} of the dimensional counting rules
by the soft  mechanism  
at the extended range of intermediate momentum transfers.
Numerically, the soft term is still more important
than the asymptotically leading contribution at rather high 
$Q^2\sim 50\div 100 GeV^2$.

Now we come back to our original problem 
of the  hard diffractive electroproduction.
Having the  experience with exclusive processes in mind,
one could expect   a similar behavior
 (i.e. a very slow approach to the asymptotically 
leading prediction)  for
  the  diffractive electroproduction as well.
The main goal of the present work is to argue that this naive 
expectation is {\bf wrong}. 
 The asymptotically leading formula starts to work already 
in the region $ Q^2 \simeq 10 \; GeV^2$, where power corrections 
do not exceed the $ 20\% $ level in the amplitude.  This 
surprising result is a consequence of very special properties 
of the WF in the $ \k $ variable (as we mentioned earlier, 
the $ \k $ dependence determines corrections to the leading term).
  At the same time, as  we argue below, these properties of WF 
are very sensitive to the QCD vacuum structure, more exactly 
to some higher dimensional vacuum condensates. As a result, 
we find that the pre-asymptotic behavior of the $\rho$-meson 
diffractive electroproduction amplitude essentially depends on the (non-) factorizability of particular mixed quark-gluon condensates. In 
our opinion, this makes diffractive processes an excellent laboratory for testing our current understanding of the QCD vacuum structure. 
In contrast to the standard QCD sum rules for static properties of hadrons \cite{SVZ} which probe most crude characteristics of the QCD 
vacuum like $ \la \bar{\psi} \psi \ra \; , \; \la g^2 G^2 \ra $, 
diffractive amplitudes are remarkably influenced by far more silent 
vacuum features. Moreover, in the latter case all these theoretical 
ideas can in principle be given a dynamical test by varying the photon virtuality $ Q^2 $ and measuring a deviation from the asymptotic 
predictions of Ref.\cite{BFGMS} (provided, of course, all other 
sources of the $ Q^2 $ dependence such as shadowing corrections 
\cite{Maor} and the $ Q^2 $ evolution of the gluon distribution 
are properly taken into account). One can therefore say that by 
studying the pre-asymptotic behavior of diffractive amplitudes
 we actually study the fine structure of the QCD vacuum.

Our presentation is organized as follows. In Sect.2 we collect relevant definitions and discuss general constraints on the nonperturbative $\rho$-meson 
WF stemming from general principles of QCD. Sect.3 deals with the calculation of the lowest moments $ \la \k \ra $ and $ \la \vec{k}_{\perp}^4 \ra $ of the $\rho$-meson transverse momentum distribution from the equations of motion and QCD sum rules.The main observation of this section is a presence of strong fluctuations in the transverse direction
 $ \la \vec{k}_{\perp}^4 \ra / \la \k \ra^2 \gg 1 $. We show how
 this ratio is fixed by particular mixed quark-gluon condensates.  
In Sect.4  we propose a model WF satisfying the general constraints
 of Sect. 2 as well as the lowest moments  $ \la \k \ra $ and $ \la \vec{k}_{\perp}^4 \ra $ of Sect.3. All these results are applied to 
the study of pre-asymptotic effects due to the "intrinsic" transverse
 momentum (the Fermi motion) in the hard diffractive electroproduction in Sect.5, where we also present our understanding of a fundamental difference between exclusive and diffractive processes.  As a by-product of our
 study, we propose a simple reformulation of the diffractive amplitude 
in terms of a current Green function.  Appendix A establishes a connection between the transverse momentum distribution we are dealing with and higher twist distribution amplitudes of  the standard ligh-cone OPE approach.
 Methods for estimates of relevant higher dimensional vacuum condensates are described in Appendix B.  Sect.6 presents a summary of our results.

\section{ General constraints on the nonperturbative wave function $\wf $.}

The aim of  this section is to provide necessary definitions 
and establish some essentially model independent constraints on the non-perturbative  
 WF  $\wf $ which follow from the use of such general methods as dispersion relations, duality and large order perturbation theory.
  For definiteness, we will consider the longitudinally polarized 
charged $ \rho$-meson. 
 
Let us start with the definition of the non-perturbative WF $ \Psi(\xi,b^2) $
in the so-called impact parameter representation as an infinite series of local 
gauge invariant operators (a low normalization point $ \mu^2 \sim 1 \; GeV^2 $
is implied and $ b^2 \equiv -z^2 $)
\bea
\label{d}
\sum_{n} \frac{i^n}{n!} \lo \bar{d}(0) \gmmu  (iz_{\nu} \nableftright_{\nu}  )^{n} u(0) \rro
 &=& f_{\rho} m_{\rho} \emu \psi(zq,b^{2})  \nonumber \\
 &=& f_{\rho} m_{\rho} \emu \int_{-1}^{1}
d \xi e^{-i \xi q z} \Psi(\xi , b^2)
\eea
Here $ \nableftright_{\nu} = \nabright_{\nu} - \nableft_{\nu} \; , \; \nabright_{\nu} = \stackrel{\rightarrow}{\partial}_{\nu} - i g A_{\nu} $ and $ \emu $ 
stands for the polarization vector.

In the asymptotic limit of hard exclusive processes \cite{Brod,Brod1} the quark and antiquark are
produced at small distances $ z \sim 1/Q \rightarrow 0 $ where $ Q $ is a typical large momentum transfer. In this case the only remaining dependence on 
the variable $ zq \sim 1 $ is given by the leading twist distribution amplitude (DA) $ \phi(\xi) $ :
\bea
\label{1}
 \lo \bar{d}(0) \gmmu  (iz_{\nu} \nableftright_{\nu}  )^{n} u(0) \rro
&=&  f_{\rho} m_{\rho} \emu (zq)^n \langle \xi^n \rangle \equiv  f_{\rho} m_{\rho} \emu (zq)^n  \int_{-1}^{1} d  \xi \xi^n \phi  (\xi)  \nonumber \\
\xi &=& 2u -1 \; ,\;  z^2 = 0
\eea
In the infinite momentum frame (IMF) $ q_{z} \rightarrow \infty $ the DA 
$ \phi(\xi)  $ describes the distribution of the total longitudinal momentum $ 
q_{z} $ between the quark and antiquark carrying the momenta $ u q_{z} $ and 
$ (1-u) q_{z} $ , respectively. In what follows we will use the 
both variables $ 
\xi $ and $ u\; , \;  \bar{u} \equiv  1-u $ interchangeously. 

 A portion of model independent  information on  the DA $ \phi(\xi) $ can be obtained using the dispersion relations and quark-hadron duality. To this end,
we study the asymptotic behavior of the correlation function
\beq
\label{2} 
i  \int dx e^{iqx}\la 0|T J_{n}^{\|}(x),J_0(0) |0\ra=
(zq)^{n+2}I_{n}(q^2),~~
 J_{n}^{\|}=\bar{d}\gamma_{\nu}z_{\nu}
(i\nableftright_{\mu}z_{\mu})^{n} u
 \eeq
(here the label $ \parallel $ stands for the longitudinal direction which is selected by the projector $ z_{\mu} \; :  z^2 = 0 \; , z_{\mu } q_{\mu} 
\neq 0 $ .)
At large $ q^2 \rightarrow  - \infty $ the exact answer for the correlation function (\ref{2} ) is given by the perturbative one-loop diagram 
\beq 
\label{3}
\frac{1}{\pi}\int_0^{\infty} ds\frac{Im I_n^{pert}(s)}{s-q^2} \; ,~~
 Im I_{n}^{pert}(s) =\frac{3}{4\pi(n+1)(n+3)}.
 \eeq
On  the other hand, in the IMF the correlation function (\ref{2}) can be parametrized in terms of intermediate hadron states, the lowest one being the 
$ \rho -$ meson. Therefore the following relation takes place 
\beq
\label{3a}
 \frac{1}{\pi}\int_0^{S_{\rho}^n} ds Im I(s)^{pert}_{n}=
\frac{1}{\pi}\int_0^{\infty} ds Im I(s)^{\rho}_{n},
 \eeq
The only assumption made here is that the $ \rho $-meson  gives a non-zero contribution in (\ref{3a}),  i.e. the duality interval $ S_{\rho}^n  ( \parallel) $ ( the subscript $ \parallel $ specifies the longitudinal direction ) does not vanish for arbitrary n. No further assumptions on highly excited states (like those imposed in the QCD sum rules approach) are needed. In this case Eq. (\ref{3a}) yields 
\beq
\label{4}
  f_{\rho}^2\la \xi^n\ra (n\rightarrow\infty)
\rightarrow\frac{3S_{\rho}^{\infty}(\|)}{4\pi^2n^2}
 \eeq
( we have used the fact that for the longitudinally polarized 
$ \rho $-meson in  the IMF $ \emu  =  q_{\mu} /m_{\rho} + O(1/q_{z}) $ .)  This 
formula unambiguously fixes the end-point behavior of the DA \cite{CZ}  :
\beq
\label{5}
 \la\xi^n\ra=\int_{-1}^1d\xi \xi^n\phi(\xi)\sim 1/n^2,~~~~~
   \phi(\xi\rightarrow
\pm 1)\rightarrow (1-\xi^2).
\eeq
We want to emphasize that the constraint (\ref{5}) is of very general 
origin
and follows directly from QCD. No numerical approximations were involved 
in the above derivation. Pre-asymptotic as $ q^2  \rightarrow -  \infty $ perturbative 
and non-perturbative corrections are only able to  change the duality  
interval in Eq. (\ref{4}) (which is an irrelevant issue, anyhow) but not 
 the parametric $ 1/n^2 $ behavior. 

The same analysis can be done for the transverse distribution. It moments are 
defined analogously to Eq.(\ref{1}) through gauge invariant matrix elements
\beq
\label{6}
\la 0|\bar{d}\gmmu
(i \nabright_{\nu}
 t_{\nu})^{2n} u|\rho(q)\ra=f_{\rho} m_{\rho} \emu 
 (-t^2)^n\frac{(2n-1)!!}{(2n)!!}\la \vec{k}_{\perp}^{2n} \ra  \;  , 
\eeq
where  transverse vector $t_{\mu}=(0,\vec{t},0)$ is perpendicular
 to the hadron momentum $q_{\mu}=(q_0,0_{\perp},q_z)$.
The factor $\frac{(2n-1)!!}{(2n)!!}$ is introduced   to
(\ref{6}) to take into account
  the integration over $\phi$ angle in the transverse plane:
$\int d\phi (\cos\phi)^{2n}/ \int d\phi= {(2n-1)!!}/{(2n)!!}$.
By  analogy with a non-gauge theory we call  $\la \vec{k}_{\perp}^{2} \ra$ in this equation
  the mean value of the quark perpendicular momentum, though  it does not have 
a two-particle interpretation. The usefulness of this object and its relation to the  higher Fock components will be discussed at the end of this section. Here we note that Eq. (\ref{6}) is the only possible way to define
 $\la \vec{k}_{\perp}^{2} \ra $ in a manner consistent with gauge invariance and 
operator product expansion.

To find the behavior  $\la \vec{k}_{\perp}^{2n} \ra $ at large $ n $ we can repeat the previous duality arguments with the following result
\footnote{ Here
and in what follows we ignore any mild (nonfactorial) $n$-dependence.}:
\beq
\label{8}
f_{\rho}^2\la \vec{k}_{\perp}^{2n} \ra \frac{(2n-1)!!}{(2n)!!}\sim
n!\Rightarrow
 f_{\rho}^2\la \vec{k}_{\perp}^{2n} \ra \sim n!
\eeq
 This behavior (for the case of pion) has been obtained in Ref.\cite{Zhit2}
by the study of large order perturbative series  for a proper 
correlation function. Dispersion relations 
and duality arguments transform this information into Eq.(\ref{8}). 
 It is important to stress that  
any nonperturbative wave function should respect  Eq.(\ref{8}) 
 in spite of the fact that
apparently we calculate only  the perturbative part
( see the
comment after Eq.(\ref{5})). The duality turn this perturbative information into exact properties of the nonperturbative WF.

The most essential feature of Eq. (\ref{8}) is its finiteness for arbitrary $ n $. This means that higher moments 
\beq
\label{mom}
\la \vec{k}_{\perp}^{2n} \ra =
\int d\k d\xi
\vec{k}_{\perp}^{2n}\Psi(\k, \xi ) 
\eeq
{\bf do exist} for any $ n $. 
In this formula we introduced the nonperturbative $ \wf $ normalized to one \footnote{To avoid possible misunderstanding, we stress that our definitions are very different from those made in the light-cone perturbation theory \cite{Brod1} where $ \la \k \ra_{BS} = \int_{-1}^{1} d \xi \int \; d^2 \vec{k}_{\perp} \vec{k}_{\perp}^{2} | \Psi_{BS}( \xi, \vec{k}_{\perp}^2) |^2 $ (here BS stands for bound state). Our approach does not suppose at all the existence of a bound state equation in QCD, but rather relies entirely on the logic of OPE.}.
Its moments are determined by the local
matrix elements (\ref{6}). The relation to the longitudinal distribution
amplitude $\phi(\xi)$  looks as follows :
\beq
\label{9}
\int d\k \wf
= \phi(\xi)  \; , \; \int_{-1}^1 d\xi\phi(\xi)=1
\eeq
The  existence of the arbitrary high moments
$\la \vec{k}_{\perp}^{2n} \ra$ means that the nonperturbative
$ \wf $ falls off at large transverse momentum $\k$
faster than any power function.
   The relation (\ref{8})
fixes the asymptotic behavior of $ \wf$ at large $\k$.
Thus, we arrive at the following constraint:
\beq
\label{10}
   \la \vec{k}_{\perp}^{2n} \ra =
\int d\k d \xi
\vec{k}_{\perp}^{2n}\Psi(\k, \xi )\sim n!  \; \; ,
 ~~~~~n\rightarrow\infty 
\eeq

We can now repeat our duality arguments again for an arbitrary number of
transverse derivatives and large ($n\rightarrow\infty$) number
of longitudinal derivatives. The result reads \cite{Zh94}:

\beq      
\label{11}   
  \int d\k
\vec{k}_{\perp}^{2k}\Psi(\k, \xi\rightarrow\pm 1 )\sim  (1-\xi^2)^{k+1}
\eeq

The constraint (\ref{11}) 
 is extremely important and implies that the $\k$
dependence of   $\Psi(\k, \xi )$ comes
{\bf exclusively in the combination}
$ \k/(1-\xi^2)$ at $\xi\rightarrow\pm 1$.  This means that the standard assumption
on factorizability  $\Psi(\k, \xi ) =\psi(\k )\phi(\xi)$
 is at variance with very general properties of the theory such as duality and 
dispersion relations. The only form of $ \wf $ satisfying all the constraints 
(\ref{5}),(\ref{10}) and (\ref{11}) is the Gaussian with a very particular argument :
\beq
\label{12}
 \Psi(\k\rightarrow\infty, \xi)\sim \exp \left(-\frac{\k}{\Lambda^2 (1-\xi^2)} \right)
 \eeq
(here $ \Lambda^2 $ is a mass scale which can be fixed by calculating the moments $ \la \k \ra , \la \vec{k}_{\perp}^4 \ra $ etc.) Strictly speaking, so far we have only established the validity of Eq.(\ref{12}) in a vicinity of the end-point region $ \xi \rightarrow \pm 1 $. In Appendix A we discuss in what sense Eq.(\ref{12}) can be approximately valid in the whole range of the $ \xi $ variable.

Here a few comments are in order. Our result (\ref{12})
essentially supports the old quark model-inspired idea on the SU(6) invariance 
of hadron WF's. Indeed, the very same analysis carries without any changes for 
pseudoscalar $ \pi $ - and $ K $ -mesons (non-zero s-quark mass effects are 
subleading). A possible violation of SU(6) is expected via a variation of the parameter $ \Lambda^2 $ for different hadrons. Moreover, the ansatz (\ref{12}) coincides with the well known harmonic oscillator WF, except for the mass term. One should stress that there 
is no room for such term in QCD. Its inclusion violates the duality constraint
(\ref{5}) since in this case we would have
\beq
\la \xi^n \ra  \sim \int_{-1}^{1} d \xi \xi^n \exp \left(- \frac{m^2}{\Lambda^2 (1-\xi^2) } \right)\sim
\exp(-\sqrt{n} ) \; , ~n \rightarrow \infty 
\eeq
instead of the $ 1/n^2 $ behavior (\ref{5}). In other words, a true nonperturbative WF must respect the asymptotic freedom which is incompatible with the quark model -type mass term in the WF \footnote{To set this more accurately, one can say that a possible (scale-dependent) mass term in $ \wf $ must renormalize to zero at a normalization point $ \mu^2 \sim a \; few \; GeV^2 $ where the duality arguments apply. This conclusion is not at variance with popular models for the QCD vacuum such as e.g. the instanton vacuum (see \cite{Shur} for review).}. This difference leads to interesting phenomenological consequences, see Ref.\cite{ChibZh} for more detail.
Furthermore, we would like to clarify the correspondence between our results (\ref{12}) and those obtained in Ref. \cite{Radtr} on similar grounds. There, it was also claimed that $ \k $ and $ \xi$ enter the WF only in the combination
$ \k/(1 -\xi^2) $  but a different WF was suggested, $ \wf \sim \Theta ( s_{0}
- \k/(1 -\xi^2) ) $. Our approach differs from that of Ref. \cite{Radtr} in that
we apply the duality arguments for currents with derivatives and in addition use  the large order perturbation theory 
analysis. As a result, we end up with the Gaussian instead of the $ \Theta $ -function.  

Our final comment concerns with the meaning of Eq.(\ref{6}). As has been explained above,  $ \la \k \ra $ is {\it not} literally
an interquark transverse momentum. The notion of interquark transverse distance 
is senseless in a gauge theory because quark transverse degrees of freedom 
are undistinguishable from longitudinal degrees of freedom of higher Fock states at any normalization point due to equations of motion, see Ref.\cite{BF} and  Appendix A. Thus an attempt to separate them would 
be at variance with exact equations of motion. Any theoretically consistent calculation of higher twist effects must include these 
two effects simultaneously at each given power of $ Q^2 $. However, the two contributions may well differ numerically. In fact, in all known examples where 
higher twist effects for exclusive amplitudes were calculated \cite{CZ,BH} a contribution 
coming from the $ \bar{\psi} G \psi $ Fock state turns out smaller than a two-particle contribution due to the transverse motion by typically a factor of 2-3 , and 
comes with an opposite sign. It is probably worth noting that in the light-cone
QCD sum rule approach to the pion form factor at intermediate $ Q^2 $ a two-particle DA of twist 4 yields the same $ 1/Q^4 $ behavior as the leading twist   DA, while a contribution due to $  \bar{\psi} G \psi $ DA is down by the extra 
power of $ Q^2 $ \cite{BH}. Thus, one can expect that a calculation based 
on the use of the light-cone WF $ \wf $ alone can serve as a reasonable estimate of 
higher twist effects. It is just the line of reasoning taken in this paper. 
In no sense we claim that a WF like (\ref{12}) exhausts the $ \rho $-meson properties,
nor we pretend to give a complete evaluation of higher twist effects in 
diffractive electroproduction (see Sect.5).

\section{ Lowest moments $ \la \k \ra  \; , \;   \la \vec{k}_{\perp}^4 \ra $ and vacuum  condensates}

The general constraints of the previous section are insufficient for building up a realistic nonperturbative WF for the $\rho$-meson. To fix the latter, we 
follow the same logic as in the analysis of the distribution amplitudes and 
calculate the lowests $ \la \k \ra $ and  $  \la \vec{k}_{\perp}^4 \ra $ moments of $ \wf $. The physical meaning of the second moment  $ \la \k \ra $ is clear:
this quantity serves as a common scale for power corrections in physical amplitudes, cf. the next section. The fourth moment  $  \la \vec{k}_{\perp}^4 \ra $ produces a next-to-leading power correction and signals on how strongly $ \wf $ fluctuates in the $ \k $ plane. 
In this section we will 
calculate the lowest moments $ \la \k \ra $ and  $  \la \vec{k}_{\perp}^4 \ra $ by a combined use of the equations of motion and QCD sum rules technique. With the 
knowledge of  $ \la \k \ra $ and $ \la \vec{k}_{\perp}^4 \ra $ we then construct a model WF $ \wf $ which will be used to estimate higher twist effects in diffractive electroproduction in Sect.5.

We start with a general Lorentz decomposition of the matrix element 
\bea
\label{26}
\frac{1}{2} \sum_{perm} \lo \bar{d} \gmmu i \nabright_{\alpha} i \nabright_{\beta} u \rro = \emu q_{\alpha} q_{\beta} \fro \mro \la u^2 \ra \nonumber \\
-\frac{1}{2} \emu \delta_{\alpha \beta}  \fro \mro   \la \k \ra + (\ealpha \delta_{\mu \beta} + 
\ebeta \delta_{\mu \alpha} ) \mro^2 A_{1} \nonumber \\
+ (\ealpha q_{\beta} q_{\mu} + \ebeta q_{\alpha} q_{\mu} - 2 \emu q_{\alpha}
q_{\beta} ) A_{2}
\eea
(here $ perm $ stands for the symmetrization in respect to $ \alpha, \beta $).  We have defined the last term such that to reproduce the usual definition 
of the first term as the second moment of the leading twist DA $ \phi(u) $ when all Lorentz indices are 
contracted with the light-like vector $ z_{\mu} $. The second term gives rise 
to the definition (\ref{6}) after contraction with the transverse vector $ t_{\mu} $. With the definition (\ref{26}) at hand, our strategy is to relate the unknown parameters $ \la \k \ra \; , \; A_{1} \; , \; A_{2} $ to some matrix elements of gauge invariant operators. At the second step these matrix elements are estimated using the QCD sum rules.
  
The first relation between the coefficients
$ A_{1}, A_{2} $ is obtained by multiplying Eq.(\ref{26})
by $ \delta_{\alpha \mu} $:
\beq
\label{27}
 \frac{\fro}{ 2 \mro} \la \k \ra - 5 A_{1} - A_{2} = 0 
\eeq
Another relation can be derived when one multiplies Eq.(\ref{26}) by $ q_{\mu} $. This yields the following constraint
\bea
\label{28}
(\varepsilon_{\alpha}^{(\lambda)} q_{\beta} + \varepsilon_{\beta}^{(\lambda)} q_{\alpha} ) \mro^2 (A_{1} + A_{2}) &=& - \lo \bar{d} \gmmu (g G_{\mu \alpha} \nabright_{\beta} + g G_{\mu \beta}  \nabright_{\alpha} ) u \rro \nonumber \\
& \equiv & - (\varepsilon_{\alpha}^{(\lambda)} q_{\beta} + \varepsilon_{\beta}^{(\lambda)} q_{\alpha} ) \mro^2 \eta 
\eea 
The matrix element (\ref{28}) parametrized by the number $ \eta $ can be estimated by studying  the following correlation function
\beq 
\label{28a}
i \int dx \; e^{iqx} \lo T \{ \bar{u} \gamma_{\rho} d(x) \; \bar{d} \gmmu (g G_{\mu \alpha} \nabright_{\beta} + g G_{\mu \beta}  \nabright_{\alpha} ) u (0)  \} \ro
\eeq
The matrix element of interest appears as the lowest intermediate hadron state contribution to the imaginary part of (\ref{28a}). On the other hand,  the correlation function (\ref{28a}) can be calculated for  $ q^2 \rightarrow - \infty $ in QCD. The perturbative contribution can be easily read off a very similar sum rule in Ref. \cite{BF} and numerically turns out negligible. The first power correction $ \sim \la g^2 G^2 \ra $  in (\ref{28a}) vanishes, and  we end up with the following simple estimate
\beq
\label{28b}
\eta \simeq \frac{8 \pi}{27} \alpha_{s} \frac{ \la \bar{\psi} \psi \ra^2}{ \fro
\mro^3 } \simeq 1.8 \times 10^{-3} \; GeV^2 
\eeq
We next contract Eq.(\ref{26}) with $ \delta_{\alpha \beta} $. Then the matrix element in the LHS becomes
\beq
\label{29}
\lo \bar{d} \gmmu (i \nabright_{\alpha})^2 u \rro = - \lo \bar{d} g \tilde{G}_{
\mu \nu} \gmnu \gmf u \rro \equiv - \emu \mro^3 \lambda
\eeq
The number $ \lambda = 24 \pm 3 \; MeV $ was calculated a long time ago \cite{BrKol} by the QCD sum rules method.
Using (\ref{27}), (\ref{28}), (\ref{28b}) we finally obtain 
\beq
\label{30}
\la \k \ra = \frac{2 \mro}{3 \fro} [ \mro \lambda + \fro \mro \la u^2 \ra
+ 3 \eta ] \simeq ( 420 \; MeV)^2
\eeq
Note that the last term in Eq.(\ref{30}) is rather small in comparison with the other two terms  (for the second term we used here the asymptotic DA $ \phi(u) = 6 u 
(1-u) $ with $ \la u^2 \ra = \int du u^2 \phi(u) = 0.3 $ ). This fact is quite understandable as the matrix element in Eq.(\ref{28}) is proportional to first moments of three-particle $ \bar{q} G q $  DA's while their normalizations are of the order of $ \mro \lambda $. Therefore, an uncertainty in Eq.(\ref{28b}) does not play an important role in our estimate (\ref{30}).
 We will use this observation in the calculation of $ \la \vec{k}_{\perp}^4 \ra $. Numerically $ \la \k \ra $ for the $\rho-$meson  (\ref{30}) is somewhat larger than the analogous parameter for the pion  $ \la \k \ra_{\pi} 
\simeq (330 \; MeV)^2 $ \cite{CZ} ( see also Eq.(\ref{25}) in Appendix A).

The same type of analysis can be done for calculation of the fourth moment 
$ \la \vec{k}_{\perp}^4 \ra $ though the algebra in this case becomes somewhat
more complicated. We write
\bea
\label{31}
\frac{1}{4!} \sum_{perm} \lo \bar{d} \gmmu i \nabright_{\alpha} i \nabright_{\beta} i \nabright_{\lambda} i \nabright_{\xi} u \rro = \emu [
q_{\alpha} q_{\beta} q_{\lambda} q_{\xi} \fro \mro \la u^4 \ra  \nonumber \\
+ (\delta_{\alpha \beta} \delta_{\lambda \xi} + perm ) \frac{1}{8} \fro \mro 
\la  \vec{k}_{\perp}^4 \ra  
+ ( \delta_{\alpha \beta} q_{\lambda} q_{\xi} + perm)
C_{1} ]  \nonumber \\ 
+ ( q_{\mu} \ealpha q_{\beta} q_{\lambda} q_{\xi} + perm - 4 \emu
q_{\alpha} q_{\beta} q_{\lambda} q_{\xi} ) C_{2} + q_{\mu} ( \ealpha q_{\beta} 
\delta_{\lambda \xi} + perm) C_{3}  \\
+ ( \delta_{\mu \alpha} \ebeta q_{\lambda} q_{\xi} + perm ) C_{4} + ( \delta_{\mu
\alpha} \delta_{\beta \lambda} \varepsilon_{\xi}^{(\lambda)} + perm) \mro^2 C_{5}
\nonumber 
\eea
Proceeding analogously to the previous case and neglecting numerically small contributions 
due to second moments of three-particle DA's, we can obtain a system of linear 
equations on the coefficients $ C_{i} $. The solution reads
\bea
\label{32}
C_{1} &=& -\fro \left( \frac{1}{5} \mro^3 \la u^4 \ra + \frac{1}{24} \frac{
 \la \vec{k}_{\perp}^4 \ra }{\mro} \right)  \nonumber \\
C_{2} &=& - \frac{3}{20} \fro \mro^3 \la u^4 \ra \nonumber \\
C_{3} &=& \frac{1}{48} \frac{\fro}{\mro}  \la \vec{k}_{\perp}^4 \ra \\
C_{4} &=& \frac{1}{20} \fro \mro^3 \la u^4 \ra \nonumber \\
C_{5} &=& -  \frac{1}{48} \frac{\fro}{\mro}  \la \vec{k}_{\perp}^4 \ra \nonumber
\eea
We finally contract Eq.(\ref{31}) with $ \delta_{\alpha \beta} \delta_{\lambda
\xi} $ and make use of (\ref{32}). Then the following equation holds
\beq
\label{33}
\frac{1}{4} \lo \bar{d} \gmmu (g \sigma G)^2 u \rro + \frac{1}{2} \lo \bar{d}
\gmmu g^2 G_{\alpha \beta}^2 u \rro  = \emu \mro^2 \left[ -\frac{3}{5}  \fro \mro^3 \la u^4 \ra + 2  \frac{\fro}{\mro}  \la \vec{k}_{\perp}^4 \ra  \right]
\eeq
We have thus reduced the calculation of the fourth moment $  \la \vec{k}_{\perp}^4 \ra $ to the calculation of the matrix elements of operators    containing only the gluon tensor $ G_{\mu \nu } $ and not the gluon potential 
$ A_{\mu} $ itself as in the original problem (\ref{6}). To estimate these  matrix elements, it proves convenient \cite{Zh94} to consider
the following nondiagonal correlation functions
\bea
\label{34}
T_{1}(p)  &=& i \int \; dx e^{ipx} \lo \bar{u} \sigma_{\alpha \beta} d(x)
 \; \bar{d} \gmmu g^2 G_{\lambda \xi}^2 u(0) \ro  \nonumber \\
T_{2}(p) &=& i \int \; dx e^{ipx} \lo \bar{u} \sigma_{\alpha \beta} d(x)
\; \bar{d} \gmmu  u(0) \ro 
\eea
Perturbative contributions vanish in the chiral limit in the both correlation functions (\ref{34})
due to an odd number of gamma matrices. Thus, a leading as $ p^2 
\rightarrow - 
\infty  $ behavior comes from condensate terms in the OPE for 
$ T_{1} , T_{2} $.
Dividing $ T_{1} $ by $ T_{2} $ we get rid of the $\rho $-meson 
residue $ \sim  \lo \bar{u}  \sigma_{\alpha \beta} d \rro $ in 
(\ref{34}). Then for  the reduced matrix elements $ \lambda_{1}, 
\lambda_{2} $ defined by 
\beq
\label{35}
 \lo \bar{d}
\gmmu g^2 G_{\alpha \beta}^2 u \rro = \emu \mro^5 \lambda_{1} \; , \;  
 \lo \bar{d} \gmmu (g \sigma G)^2 u \rro  = \emu \mro^5 \lambda_{2}
\eeq
we obtain after a simple algebra the following relations for the 
matrix elements of interest in terms of mixed vacuum condensates of the dimension 7 :
\bea
\label{36}
\lambda_{1} &\simeq & \frac{\fro}{\mro^4} \frac{ \lo \bar{u}
 (gG)^2 u \ro }{ \lo
\bar{u} u \ro }   \nonumber \\
\lambda_{2} &\simeq & \frac{\fro}{\mro^4} \frac{ \lo \bar{u}
 (g\sigma G)^2 u \ro }{ \lo
\bar{u} u \ro }   
\eea
Here the second of Eqs.(\ref{36}) is obtained analogously to (\ref{34}).
Using Eq.(\ref{33}) we finally get 
\beq
\label{37a}
 \la \vec{k}_{\perp}^4 \ra \simeq \frac{3}{10} \mro^4 \la u^4 \ra +
 \frac{1}{4}
 \frac{ \lo \bar{u} (gG)^2 u \ro }{ \lo
\bar{u} u \ro } + \frac{1}{8} \frac{ \lo \bar{u} (g\sigma G)^2 u \ro }
{ \lo
\bar{u} u \ro }
\eeq
This is the main result of this section: we have explicitly expressed
 $  \la \vec{k}_{\perp}^4 \ra $ in terms of the vacuum expectation values (VEV's) of the dimension 7 operators. Naively, one could estimate these condensates by factorizing them into the products of the quark
 $ \la \bar{\psi} \psi \ra $  and gluon $ \la g^2 G^2 \ra $ condensates. 
This procedure, based on the factorization hypothesis, does not work 
in the given case: there are essential deviations from the factorization prediction in VEV's (\ref{36}).  The non-factoraziblity of mixed quark-gluon matrix 
elements of such type has been studied in \cite{Zh93,Zh94} by two independent methods with full agreement in estimates between them. The first one
 \cite{Zh93} was based on the analysis of heavy-light
quark systems which allows to obtain restrictions on the VEV's like 
(\ref{36}),
while the second method has related the vacuum condensates of the form (\ref{36}) to some pion matrix elements known from PCAC. As our results essentially depend on the magnitude of the condensates (\ref{36}), we 
shortly review the method of Ref.\cite{Zh94} in Appendix B. Here we only 
formulate the result of this analysis. A measure of  non-factorizability introduced by the correction factors $ K_{1} \;  , \; K_{2} $ in the matrix elements
\bea
\label{37b}
  \lo \bar{u} (gG)^2 u \ro &=& \frac{1}{6} K_{1} \lo g^2 G^2 \ro \;  \lo  
\bar{u} u \ro   \nonumber \\
\lo \bar{u} (g\sigma G)^2 u \ro &=& \frac{1}{3} K_{2}  \lo g^2 G^2 \ro \;  \lo
\bar{u} u \ro 
\eea
( $ K_{1} = K_{2} = 1 $ in the factorization limit)  is 
approximately the same $ K_{1} = K_{2} =  K \simeq 3 $ for the both mixed operators appearing in Eq.(\ref{36}). A possible uncertainty of this estimate does not exceed 30 \%  \cite{Zh94}. 
Using Eqs.(\ref{37a}) and (\ref{37b}) we arrive at the following numerical  estimate
\beq
\label{37}
 \la \vec{k}_{\perp}^4 \ra \simeq \frac{1}{8} \left[ \frac{K}{3}  \la g^2 G^2 \ra  +  \frac{K}{3}  \la g^2 G^2 \ra  + \frac{12}{5} \mro^4 \la u^4 \ra
\right] \simeq 0.14 \; GeV^4
\eeq
This large value of  $ \la \vec{k}_{\perp}^4 \ra $ seems rather surprising as 
it implies strong fluctuations of the nonperturbative WF $ \wf $. To measure the magnitude of these fluctuations it is convenient to introduce a dimensionless parameter 
\beq
\label{38}
R \equiv \frac{  \la \vec{k}_{\perp}^4 \ra}{ \la \k \ra^2} \simeq
       4.2   \;   \mbox{ if  $ \la \k \ra \simeq (420 \; MeV)^2 $}     
\eeq
The phenomenon of strong fluctuations in $ \wf $ has been found previously by analogous 
methods for the pion wave function in Ref. \cite{Zh94} with approximately the same number for the ratio (\ref{38}). In terms of the QCD vacuum structure these fluctuations are due to the numerical enhancement of the high dimensional quark gluon mixed condensates or, what is the same, the large magnitude of the parameter K in Eq.(\ref{37b}).  

\section{Model wave function $ \wfu $ }

The results obtained so far are essentially model independent. We have fixed 
the form of the high-$\k $ tail of the true nonperturbative WF $ \wf $ Eq.(\ref{12}) by the use of the quark-hadron duality and dispersion relations. Furthermore, we calculated the lowest moments $ \la \k \ra $ and $  \la \vec{k}_{\perp}^4 \ra $ using the equations of motion and QCD sum rules.
Now our purpose is to build some model for the true nonperturbative WF $ \wf $ 
which would respect all general constraints of Sect.2 and incorporate the effect of strong fluctuations in the transverse $ \k $ plane found in the previous Sect.3 (see Eq.(\ref{38})). In spite of the fact 
that this inevitably brings some model dependence into any physical quantity calculated within such a wave function, there remains a region where the results of such 
a calculation become practically model independent. This happens when the 
answer is essentially determined only by the lowest $  \la \k \ra $ and $  \la \vec{k}_{\perp}^4 \ra $ but not higher $ \la \vec{k}_{\perp}^6 \ra $, etc. 
moments of the WF. As we shall see in Sect. 5, this is precisely the case in the hard diffractive electroproduction for sufficiently high $ Q^2 > 10 \; GeV^2 $.
 
The effect of strong $ \k $ fluctuations (\ref{38}) cannot be taken into account 
by a simple Gaussian ansatz of the type (\ref{12}) which would yield the factor of 15/7 in Eq.(\ref{38}). The only possible way to satisfy Eq.(\ref{38}) is to 
have a second pick in $ \wf $ which can have a small magnitude but reside 
far away from the the first Gaussian one. To reconcile the existence of this second pick 
with the asymptotic behavior (\ref{12}) it must enter as a pre-asymptotic term which would provide the constraint (\ref{38}). Following Ref.\cite{ChibZh} we suggest 
the following ansatz  for $ \wf $ :
\beq
\label{39}
\Psi(u, \k ) = A \left\{ \exp \left[- \frac{\k}{8 \beta^2 u \bar{u}} \right]  + 
c \exp \left[- ( \frac{\k}{8 \beta^2 u \bar{u}} - l)^2 \right] \right\}
\eeq
In this ansatz the parameters $ c $ and $ l $  determine the magnitude and position of the second pick, respectively. We have found that the best fit corresponds to the choice $ \beta = 125 \; MeV \; , \; c=0.15 \; , \; l =30 $ 
. For these values of parameters $ R \simeq 4.1 \; , \;  \la \vec{k}_{\perp}^4 \ra \simeq 0.13 $. For $ u \simeq 1/2 $ the second pick is located at $ \k \simeq 1 \; GeV^2 $. The existence of two characteristic scales in $\wf$
( 0.4 GeV and 1 GeV) is rather interesting and may probably be linked to 
phenomenological models for the QCD vacuum like the instanton vacuum model \cite{Shur} and/or the constituent quark model. If one does not insist on (or disbelieves) the effect of amplified fluctuations in the transverse $ \k $ plane (\ref{38}), then the second exponent in Eq.(\ref{39}) should be omitted. It is important, however, that in order to have the same second moment $ \la \k \ra $ the parameter $ \beta $ must be simultaneously rescaled to $ \beta \simeq 330 \; MeV $. In this case the WF takes the following form :
\beq
\label{39a}
\Psi(u, \k ) = A  \exp \left[- \frac{\k}{8 \beta^2 u \bar{u}} \right] \; , \; \beta = 330 \; MeV  
\eeq
We will use in Sect.5 the both forms in order to reveal the importance of large $ \la \vec{k}_{\perp}^4 \ra $  (\ref{37}) and, respectively, of the non-factorizability of VEV's (\ref{37b}) in the pre-asymptotic behavior of the diffractive amplitude.

The model WF (\ref{39}) meets all general
and numerical constraints derived in previous sections of this paper. The integration of (\ref{39}) over $ \k $ yields the asymptotic form of the distribution amplitude $ \phi(u) = 6 u (1-u) $ which is known to be a good approximation to the true non-perturbative DA of the $\rho $-meson. Furthermore,
numerical coefficients we have obtained for the ansatz (\ref{39}) are not very 
different from those found in \cite{ChibZh} for the case of the pion WF. 
Our conclusion is that the transverse momentum distributions in pion and 
$ \rho $-meson are to a large extent alike. 

\section{Hard diffractive electroproduction}

In this section we come back to our initial aim. The results of 
proceeding 
sections expressed in condensed form by Eq.(\ref{39}) will be used
 for the study
of the pre-asymptotic effect due to the Fermi motion in diffractive electroproduction of the $\rho$-meson. Our prime goal is to get an 
estimate for 
the onset of the asymptotic regime in this problem. 

The applicability of perturbative QCD (pQCD) to the asymptotic limit 
of the hard
diffractive electroproduction of vector meson was established in Ref.\cite{BFGMS} using the light-cone perturbation  theory. The 
authors of \cite{BFGMS} have proved that for the production of 
longitudinally polarized vector mesons by longitudinally polarized
 virtual photons the cross section 
can be consistently calculated in pQCD. There was found that at 
high $ Q^2 $ the amplitude factorizes in a product of DA's of the 
vector meson and virtual 
photon, the light-cone gluon distribution function of a target, and a perturbatively calculable on-shell scattering amplitude of a 
$ q \bar{q} $ pair off the gluon field of the target. After 
the factorization of gluons from the 
target the problem seems to be tractable within OPE-like methods.

It would be desirable to have an explicit realization to the
 idea of treating
the process of interaction of the virtual photon with the gluons
 from the target and a sequent conversion into a vector meson as 
some kind of a current Green function, to which methods of the
 light-cone OPE could be applied. Such reformulation has proved
 to simplify enormously the analysis of the pion form
factor and some other exclusive amplitudes. This approach is 
particularly
convenient for the study of higher twist effects in exclusive 
processes, where 
calculations of light-cone pQCD become tremendously complicated. 
Here we 
propose such a method. Though no new result will be given, we 
believe it is still
worthwhile to present it in this paper. The reason for this is several-fold. 
First, our method  reproduces the results of Ref.\cite{BFGMS} more
 easily and 
moreover extents them by taking into account an asymmetry of the gluon distribution of the target. Second, it allows to establish a connection
 with the nonperturbative  $\rho$-meson WF defined within the OPE.
 Note in this respect that the question on what WF should be substituted 
in the amplitude is of peculiar interest and 
causes some controversy in the literature (see below on this).  Third, the approach we suggest is ideally suited for a complete evaluation of twist 4 effects including three-particle $ \bar{q}G q$ DA's (see discussion 
in Sect.2). 
Though we do not carry such a calculation, it can be done within this
 formalism 
provided the  $ \bar{q}G q$ DA's are known. Thus, our reformulation
 will be used below merely to rederive the formulae of Ref.\cite{BFGMS,FKS} within OPE. This will enable to show that the WF which enters Eq.(\ref{46}) below is in fact the soft light-cone WF (\ref{mom}) defined as a set 
of matrix elements which has been discussed in the previous sections
 of this paper.
  
Let us start with the general form of the amplitude as a matrix 
element of the 
electromagnetic current $ j_{\mu} = e ( 2/3 \bar{u} \gmmu u - 
1/3 \bar{d} \gmmu 
d ) $ :
\beq
\label{40}
M = \epsilon_{\mu} \la N(p - r) \rho (q+ r) |  j_{\mu}(q) | N(p) \ra
\; , 
\eeq
where $ \epsilon_{\mu} $ is the polarization vector of the photon 
(only the longitudinal polarization is considered, see \cite{BFGMS})
 and $ r $ stands for the momentum transfer. We will neglect the 
masses of the nucleon and 
$ \rho$-meson in comparison with the photon virtuality $ Q^2 $ : 
 $ \mro^2 = m_{N}^2 = 0 $. For the momentum transfer $ r $ we consider
 the limit  $ r^2
= 0 $, but $ r_{\mu} \neq 0 $. 
 We next note the following. The factorization 
of the gluon field of the target \cite{BFGMS}, \cite{Strikman} means 
that that these 
gluons act as an external field on highly virtual quarks produced by
 the photon
with $ Q^2 \rightarrow \infty $. This statement can be given a formal
 meaning
by introducing into the QCD Lagrangian an additional term $ \delta S
 = i g
\int dz j_{\mu}^a (z) G_{\mu}^a  $ where $ j_{\mu}^a = \bar{q} 
\gmmu t^a q $ is the color current and $ G_{\mu}^a $ stands for 
the external gluon field from the
target. Expanding the path integral to the second order in this
 external field
corresponds to the two-gluon exchange :
\beq
\label{41}
M = -g^2 \epsilon_{\mu} \int dz dy \la N(p - r) | G_{\alpha}^a (z)
G_{\beta}^b (y) | N(p) \ra \la \rho(q + r) | T \{ j_{\mu}(0)
 j_{\alpha}^a(z) j_{\beta}^b (y) \}  \ro 
\eeq
One has to stress that our Eq.(\ref{41}) is not another "proof" of factorization, but rather should be considered as a reformulation 
of results of Ref.\cite{Strikman} (and also \cite{Radasym}, see 
Eq.(\ref{44}) below) in a way which is very suitable for practical
 calculations.
Following Ref.\cite{Radasym}, we find it convenient to choose the
 light-cone gauge for the external gluons
\beq
\label{42}
G_{\mu} (z) = n^{\nu} \int_{0}^{\infty} d \sigma G_{\mu \nu}(z + 
\sigma n) \; \; , \; \; n^{\mu} G_{\mu} = 0
\eeq
in which the nucleon matrix element can be parametrized in terms 
of the asymmetric gluon distribution function \cite{Radasym} :
\bea
\label{43}
\la N(p- r) | G_{\alpha}^a (z) G_{\beta}^b (y) | N(p) \ra = \frac{ 
\delta^{ab}}{
N_{c}^2 - 1} \frac{\bar{u}(p-r) \hat{n} u(p)}{ (-4)(pn)} 
\left( g_{\alpha \beta} - \frac{ p_{\alpha} n_{\beta} + p_{\beta}
 n_{\alpha}}{ (pn)} \right) \nonumber \\
\times \int_{0}^{1} d X \frac{F_{\xi}^g (X)}{X(X-\xi + i \delta)} 
\left( e^{-iXpz
+ i(X-\xi)py} + e^{i(X-\xi)pz - i Xpy} \right)
\eea 
Here one exploits the fact that in the limit $ r^2 = 0 $ the 
momentum transfer is proportional to the nucleon momentum $ r_{\mu} =
 \xi p_{\mu} $ with $ \xi = Q^2/(2 pq)
$ being the Bjorken variable. Under this circumstance the fractions of
 the total nucleon momentum carried by the gluon are $ xp_{\mu} +
 y r_{\mu} = (x + \xi y ) p_{\mu} \equiv X p_{\mu} $ and $ (x- 
\xi \bar{y}) p_{\mu} \equiv (X - \xi)p_{\mu} $, where $ \bar{y}
 \equiv 1-y $ . It is thus convenient \cite{Radasym}   to parametrize 
the gluon distribution in terms of the asymmetric gluon distribution
 $ F_{\xi}^g (X) $ which goes over to the usual gluon distribution
 function $ X f_{g}(X) $ in the symmetric limit $ \xi \rightarrow 0 $.
 
Using the translational invariance and the fact that $ \bar{u}(p-r) 
\hat{n}
u(p)/(-4 pn) = \sqrt{1-\xi}$ , we finally present 
the amplitude in the form
\bea
\label{44}
M = - \frac{g^2}{N_{c}^2 -1} \left( g_{\alpha \beta} - \frac{ p_{\alpha} n_{\beta} + p_{\beta} n_{\alpha}}{ (pn)} \right) \epsilon_{\mu} 
\sqrt{1-\xi}
\int_{0}^{1} d X \frac{F_{\xi}^g (X)}{X(X-\xi + i \epsilon)} \nonumber \\
\times \int dz dy e^{-iqz} 
\left( 
+ e^{i(X-\xi)py} + e^{ - i Xpy} \right) \la \rho(q+r) |T \{ j_{\mu}(z) j_{\alpha}^a (0) j_{\beta}^a (y) \} | \ro 
\eea
We have thus reduced the amplitude of diffractive electroproduction to 
the 
three-point correlation functions of quark currents between the vacuum
 and the 
$ \rho$-meson states. One can see that this correlation function is 
dominated by the ligh-cone and thus a leading contribution is given by a combination of 
non-local quark -antiquark string operators of twist 2 whose matrix 
elements yield
the leading twist DA of the $\rho$-meson. Coefficients in front of these operators are due to most singular parts of quark propagators near the light-cone. Retaining only this contribution to (\ref{44}), we recover 
the 
asymptotic answer of Ref.\cite{BFGMS} in the form presented without 
derivation in 
\cite{Radasym} :
\beq
\label{45}
M = \frac{4 \pi \sqrt{2} e \alpha_{s} \fro}{N_{c}} \frac{1}{Q}
 \int_{0}^{1} 
dX \frac{F_{\xi}^g (X) \sqrt{1-\xi} }{ X(X-\xi + i \delta)} 
\int_{0}^{1}
\frac{\phi(u)}{ u \bar{u} } 
\eeq
where $ \phi(u) $ is the standard light-cone DA (\ref{1}) (redefined 
to the interval $ u = [0,1] $ ) whose asymptotic form is $ \phi^{as}
(u) = 6 u \bar{u} $. Remind here that it is defined as $ \phi(u) = 
\int d \k \wfu $ where the soft WF $ \wfu $ is introduced by
 Eq.(\ref{mom}). 
\par
 Going over to higher twist corrections to the asymptotic answer
 (\ref{45}), one has to distinguish between two classes of contributions. Corrections of the first class come from keeping less singular 
terms in the quark propagators in (\ref{44}). Relevant terms in 
the light-cone expansion
for the quark propagator contain an additional gluon which is to 
be attributed 
to the final meson and expressed in terms of three-particle DA's. A corresponding calculation is technically involved, though it is 
feasible within the formalism presented here. The second class of 
corrections is due to the quark transverse degrees of freedom (the Fermi motion). Calculating only this contribution we make an educated guess
 on the scale of higher twist corrections in the diffractive 
electroproduction (see the comment at the end of Sect.2). 
Technically, retaining this correction amounts to
the multiplication of (\ref{45}) by 
 the following correction factor  \cite{FKS} (following the notations of Ref.\cite{FKS} we reserve the symbol $ T(Q^2) $ for the correction 
in the cross section)
\beq
\label{46}
\sqrt{T(Q^2)} = Q^4 \frac{\int_{0}^{1} \frac{du}{u \bar{u}} 
\int_{0}^{Q^2} \; d^2 \vec{k}_{\perp}   \Psi(u, \vec{k}_{\perp}^2) 
\frac{1}{(Q^2 + \k/ (u \bar{u}))^2} 
\left( 1 - 2 \frac{ \k/( u \bar{u})}{ (Q^2 + \k/( u \bar{u}) } 
\right)      
}{ \int_{0}^{1} \frac{du}{u \bar{u}} \int_{0}^{Q^2} \; d^2 
\vec{k}_{\perp}  \Psi(u, \vec{k}_{\perp}^2)} 
\eeq
where $ \wfu $ was defined earlier in terms of the local matrix elements (\ref{mom}). By definition, in the asymptotics $ Q^2 \rightarrow
 \infty $  $ \sqrt{T}  = 1 $. Deviations from $ \sqrt{T} = 1 $ 
determine a region of applicability of the asymptotic formula (\ref{45}). 

Now we are in position to discuss numerical estimates for the 
correction factor 
(\ref{46}) in order to answer the main question formulated in 
Introduction.
The authors of \cite{FKS} have observed that the choice of
 $ \wfu $ in a factorized form $ \wfu = \phi(u) 
\psi( \vec{k}_{\perp}^2 ) $ leads to a very slowly 
raising function $ \sqrt{T(Q^2)} $ which approaches 0.8 at rather
 high $ Q^2 \geq 40 \; GeV^2 $ depending on the model chosen
 for $ \wfu $. On the other hand, as we stated earlier, Eq.(\ref{44}) 
unambiguously demonstrates that it is the soft nonperturbative
 WF (\ref{mom}) that 
has to be substituted in Eq.(\ref{46}). Note in this respect that any perturbative tails like $ 1/\k $ in the WF are by definition absent 
within the logic based on a use of OPE. Such contributions effectively 
reappear only as radiative corrections to leading $ O(\alpha_{s}^0 ) $ 
results. It is clear that these corrections are always small unless
 there are special reasons which make them parametrically leading as 
$ Q^2 \rightarrow \infty $ (this is the case e.g. for the pion form
 factor). Using the model WF (\ref{39}) we obtain
a very fast approach to the asymptotics in (\ref{46}). The correction 
factor $ \sqrt{T(Q^2)} $ reaches the value 0.8 already at $ 
Q^2 \simeq 10 \; GeV^2 $   
that implies a rather low onset for the asymptotic regime in 
the amplitude
(\ref{44}).

\begin{figure}
\epsfysize=3in
\epsfbox[-50 207 558 545] {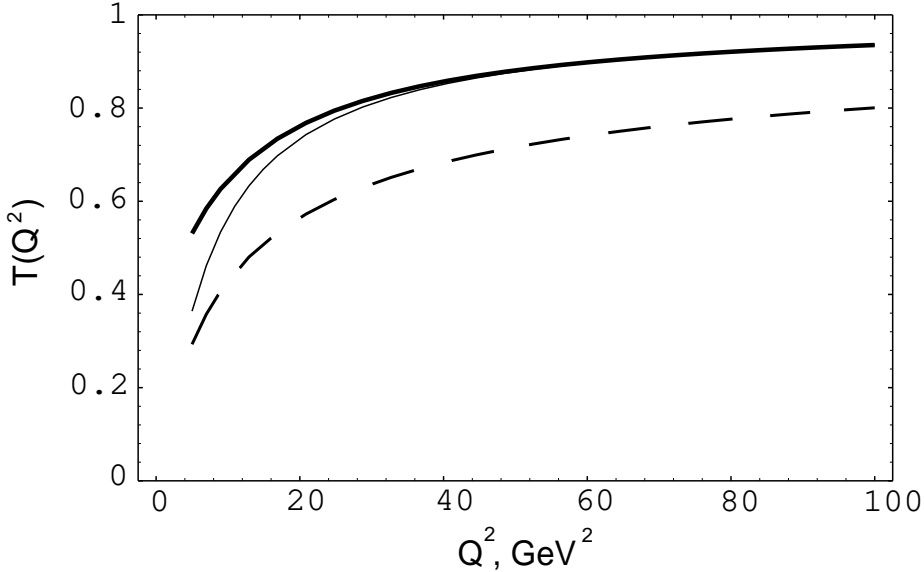}
\caption{ {\it The correction factor $ T(Q^2) $ in the cross
 section for different choices of the $ \rho$-meson WF $ \wfu $.
\label{figure} }}
\end{figure} 

 The technical reason for this behavior is quite clear: large values
 of   $ \k /(
u \bar{u}) $ are exponentially suppressed by the WF (\ref{39}) and 
thus this term remains much smaller that $ Q^2 $ for $ Q^2 > 10 \;
 GeV^2 $. The result of our calculation is shown for the correction
 $ T(Q^2) $ in the cross section on the Figure (the thick curve).
 To demonstrate the importance of amplified fluctuations in the 
$\k-$plane (\ref{38}) (and thus of the non-factorization property
 (\ref{37b}) ), we also show (the thin curve) the result of the
 calculation when the WF (\ref{39a}) not including the second pick 
is used. One sees that the difference between the two curves disappears
 for $ Q^2 > 40 \; GeV^2 $ , while it becomes quite sizeable (20 - 30 
\% ) in the region $ Q^2 = 5-10 \; GeV^2 $. One may therefore speculate 
that the non-factorizability of higher dimensional condensates 
(\ref{37b}) could be tested experimentally by measuring the cross
 section in this region. 
Still, one has to note that prior to any quantitative prediction, a 
more refined analysis 
is required in this region, including a complete evaluation of twist 4 corrections due to the $ (\bar{\psi} G \psi ) $ Fock state. For sake of comparison with the discussion of Ref.\cite{FKS} we also plot (the dashed
 curve) the factor $ T(Q^2) $ corresponding to the choice \cite{FKS}
 of the WF in the factorized "dipole" form $ \wfu \sim u \bar{u} / ( \k + \mu^2)^2 $ for $ \mu = 300 \; MeV $. Note that for such WF the second
 moment $ \la \k \ra $, as defined by Eq.(\ref{mom}), does not exist at all , and can only be introduced by means of the bound state equation, see the footnote after Eq.(\ref{mom}). The choice $ \mu = 300 \; MeV $ 
corresponds to $ \la \k \ra_{BS} \simeq (220 \; MeV)^2 $ (this number
 follows from nowhere and is taken exclusively for illustrative purposes).
What matters, however, is the fact that all other options discussed in \cite{FKS} for a factorized form of WF $ \wfu = \phi(u) \psi(\k) $ 
( including the above ansatz with $  \la \k \ra_{BS} \simeq (300 \;
 MeV)^2 $ and  $  \la \k \ra_{BS} \simeq (600 \; MeV)^2 $ and some other choices) all lie under this curve. One can therefore see quite 
unambiguously the difference in the correction $ T(Q^2) $ for the 
factorized   $ \wfu = \phi(u) \psi( \vec{k}_{\perp}^2 ) $ and 
the non-factorized 
WF (\ref{39}) in the whole range of $ Q^2 $. Furthermore, this difference is critical for the issue of the onset of asymptotic regime
 which is of our prime interest in this paper. While for WF (\ref{39}) the correction $ \sqrt{T} $  in the amplitude reaches the value 0.8 already
 at $ Q^2 \simeq 10 \; GeV^2 $ (or equivalently $ T(10 \; GeV^2) 
\simeq 0.64 $), for the "dipole" ansatz  $ \wfu \sim u \bar{u} / 
( \k + \mu^2)^2 $ with $ \mu = 300 \; MeV $ this happens only at 
$ Q^2 \simeq 35 \; GeV^2 $. For other choices of a factorized WF 
$ \wfu = \phi(u) \psi( \vec{k}_{\perp}^2 ) $ discussed in \cite{FKS} a corresponding value is pushed further to even higher $ Q^2 > 40 \;
 GeV^2 $. On the other hand, at $ Q^2 = 10 \; GeV^2 $ the difference 
between our prediction for $ T(Q^2) $ and models discussed in \cite{FKS} constitutes at least a factor of two.
  
Our use of the ansatz (\ref{39}) can, however, meet objections
as apparently it yields a large contribution to the integral (\ref{46})
 coming from
the region of small $ \k $ which is expected to be cut off on physical 
grounds. 
Here we remind that the ansatz (\ref{39}) has been 
suggested to satisfy the general constraints of Sect.2 and the 
particular values
of the moments (Sect.3). On the other hand, exclusive amplitudes 
at large $ Q^2 $
are generally only sensitive to the high-$\k $ tail of the WF which
 dominates the moments $ \la \k \ra $, $ \la \vec{k}_{\perp}^4 \ra $ etc. 
Thus one expects that these are only global 
characteristics like  $ \la \k \ra $ that matter in Eq.(\ref{46}). It 
is easy to see that this is indeed the case as soon as the WF is defined to possess all moments, see Eq.(\ref{mom}). Powers of $ 1/(u \bar{u}) $ arising when the integrand in (\ref{46}) is expanded in a series over $ \k $  are harmless due to the general constraint (\ref{12}) which excludes end-point singularities of the u-integration. In this case the $ u- $ and $ \k - $ integrations in Eq.(\ref{46})  decouple with the exponential in $ Q^2 $ (in practice this means $ Q^2 > 5 \; GeV^2 $
)
accuracy, and the answer is therefore directly expressed in terms of moments  
\beq
\label{47}
\sqrt{T(Q^2)} = 1- 20\frac{\la \k \ra}{Q^2} + 210 \frac{ \la 
\vec{k}_{\perp}^4 \ra }{ Q^4 } + \ldots = 1- \frac{3.4 \; GeV^2}{Q^2} + \frac{28.4 \; GeV^4}{Q^4} + \ldots
\eeq
Therefore our approach is self-consistent in the sense that any question on a large distance contribution to the integral (\ref{46}) is lifted up to the definitions of the moments, as has to be expected on general grounds. 
Here we emphasize again that our definitions of moments are specific to 
OPE, see the footnote after Eq.(\ref{mom}). The numbers in Eq.(\ref{47}) illustrate the observed fast growth of $ T(Q^2) $: power corrections tend to cancel each other before they become small separately. On the other hand, 
it can be seen from Eq.(\ref{47}) that for $ Q^2 < 10 \; GeV^2 $ higher
 moments of $ \wf $ turn out essential and thus in this region the 
answer for $ T(Q^2) $ is to some extent model dependent. 

Thus our final conclusion is that the onset of the asymptotic regime for 
diffractive electroproduction of the longitudinally polarized $\rho$-meson is 
approximately $ Q^2 \simeq 10 \; GeV^2 $ where corrections due to the quark transverse degree of freedom constitute less than 20 \% in the amplitude.  
This can be traced back to the fact that in the case at hand the power 
corrections are given by the matrix elements of local operators and in fact are 
fixed completely by the independent calculation of the moments. This is the 
consequence of the structure of Eq.(\ref{46}) and the fact that the WF is 
a function of the single variable  $ \k /(u\bar{u} ) $. Only global, but not local characteristics of $ \wfu $ in the $ \k $ plane are important. This situation can be 
confronted with the case of exclusive processes. In the most well studied problem of 
the pion form factor the asymptotic regime has been found to be pushed further
to $ Q^2 \gg 10 \; GeV^2 $ \cite{BH,ChibZh}, where the subleading "soft" contribution is still larger than the leading asymptotic one. One reason for this \cite{BH} is that 
a special power correction comes due to the Feynman end-point mechanism for 
the form factor which provides the whole answer $ \sim 1/Q^4 $ in absence of radiative corrections. The asymptotic answer $ \alpha_{s}(Q^2) /Q^2 $ is in fact the radiative correction to this soft term which nevertheless wins in the asymptotics $ 
Q^2 \rightarrow \infty $. In our opinion, this is precisely where the most fundamental difference between exclusive and diffractive amplitudes lies. The academic $ Q^2 \rightarrow \infty $ limit of the former is given by the hard rescattering mechanism which is $ O(\alpha_{s}) $ and thus is superseded at all 
available energies by the nonperturbative soft contribution which is down by the power $ 1/Q^2 $, but at the same time does not contain a smallness 
due to $ \alpha_{s} $. On contrary, in the diffractive amplitude power corrections which 
are again $ O(\alpha_{s}^0 ) $  ( but not attached to the end-points of the u-integration) compete with the asymptotic amplitude which itself is 
also  $ O(\alpha_{s}^0 ) $. It follows from this argument that the
 onsets of the asymptotic regime for the exclusive and diffractive 
amplitudes differ {\bf parametrically } by the factor $ 1/\alpha_{s} $. 
This conclusion is not in contradiction with the fact \cite{ChibZh}
 that for the pion form factor the nonleading soft contribution is 
larger than the asymptotic answer up to very high $ Q^2 = 50 - 100 
\; GeV^2 $.   
We are thus convinced that the analogy with the pion form factor
 problem {\bf does not }
work in the case at hand. There are no reasons to expect the onsets of the asymptotic regime to be similar in the diffractive electroproduction and exclusive processes. On contrary, they differ parametrically in  
$ 1/\alpha_{s} $. Moreover, the explicit calculations suggest that the asymptotic regime in the $ \rho$-meson diffractive electroproduction starts already at $ Q^2 \simeq 10 \; GeV^2 $. We stress that this conclusion refers only to the diffractive electroproduction of the longitudinally polarized $\rho$-meson, the situation with the transverse polarization or diffractive charmonium production can be quite different.

\section{Summary}

We hope that we have succeeded in explaining a few facts by this paper.
 Below we collect them together. \\
1. Very general principles of the theory such as the dispersion relations and quark-hadron duality allows one to fix the asymptotic form of the light-cone 
WF $ \wfu $. It is the Gaussian of the single argument
$ \wfu \sim 
\exp( - \k / 8 \beta^2 u \bar{u} ) $ , while the mass term in WF or/and the factorized WF $ \wfu = \phi(u) \psi( \k ) $ are absolutely forbidden by the above principles.  \\
2. The light-cone WF $ \wfu $ does not have a two-particle interpretation
 and is related by equations of motion to higher twist distribution 
amplitudes. However, there exists a numerical difference between their contributions, which allows to retain only the former to estimate
 an importance of higher twist effects for a particular process.    \\
3. Properties of $ \wfu $ are strongly influenced by the fine structure 
of the QCD vacuum. More precisely, the $\rho$-meson WF $ \wfu $ is very sensitive to the non-factorization of quark gluon vacuum condensates of dimension 7.  \\
4. The pre-asymptotic behavior of the $ \rho $-meson diffractive electroproduction amplitude is critically depending on properties of 
the WF $ \wfu $. When the correct WF is substituted in the amplitude, the asymptotic regime starts at rather low $ Q^2 \simeq 10 \; GeV^2 $, in drastic contrast to exclusive processes where the onset of the asymptotic regime is pushed to much higher $ Q^2 = 50 - 100 \; GeV^2 $. \\
5. This vast difference between exclusive and diffractive processes is not accidental, but rather has the clear origin : the asymptotic regime for the former is delayed in comparison to the latter parametrically by the factor $ 1/\alpha_{s}(Q^2) $. \\
6. In the intermediate energy region $ Q^2 = 5 - 10 \; GeV^2 $ the diffractive amplitude is sensitive to the non-factorization of dimension 7 vacuum condensates. Thus, the fine structure of the QCD vacuum is probed there. By preliminary estimates,
the non-factorization has an effect of the order of 20 - 30 \% in the cross section.  \\
7. To get a quantitative description of the diffractive amplitude for intermediate energies $ Q^2 = 5 - 10 \; GeV^2 $ a more refined calculation is required, including a careful analysis of a model dependence of predictions for such $ Q^2 $ and complete evaluation of twist 4 effects with account for three-particle $ \bar{q} G q $ distribution amplitudes.  \\
8. The latter technically involved calculation can be considerably simplified using 
our representation of the diffractive amplitude in terms of the current Green function, Eq.(\ref{44}) which re-expresses the results of \cite{Strikman,Radasym} in a form convenient for practical purposes. Necessary for this calculation $ \bar{q} G q $ distribution amplitudes for $\rho$-meson 
will be discussed elsewhere \cite{Hal}.   

\section*{Acknowledgments}  

We are very grateful to L. Frankfurt and M. Strikman for numerous stimulating discussions and constructive criticism. We would like to thank W. Koepf for a correspondence on the subject of Ref.\cite{FKS}. ARZ thanks the Institute for Nuclear Theory at the 
University of Washington for its hospitality and partial support
during his visit in May 1996 which initiated this study.

\clearpage
\appendix 
\def\theequation{\thesection.\arabic{equation}}

\section*{Appendix A}

\def\thesection{A}
\setcounter{equation}{0}
 
The purpose of this appendix is to  establish the connection of the problem of transverse 
momentum distribution with the standard light-cone operator product expansion 
(OPE) and explain how $ \la \k \ra $ can be directly expressed in terms of some three-particle matrix elements.  We consider this problem on the example of the pion. This choice is motivated by the fact that an analysis of higher twist distribution amplitudes in this case is considerably simplified in comparison to the $ \rho$-meson due 
to smaller number of independent Lorentz structures. The complete set of three-particle DA's of twist 4 beyond the asymptotic regime has been constructed
in Ref.\cite{BF} using the (one-loop) conformal symmetry and the QCD sum rules technique. Here we use these results to show that the combination $ b^2 u \bar{u} $ ( or $ \k/(u \bar{u} ) $ after the Fourier transform  ) is a single argument in the pion WF to the leading conformal spin accuracy. This is a justification for our choice of the model WF (\ref{39}) where the only argument $ \k / u \bar{u} $ appears.

The twist 4 three-particle DA's of interest are introduced as follows
\bea
\label{13}
\lo \bar{d}(-x) \gamma_{\alpha} \gmf  g G_{\mu \nu} (vx) u(x) | \pi(q) \ra =
 q_{\alpha}(  q_{\mu} x_{\nu}- q_{\nu} x_{\mu}) \frac{1}{qx} \int D (\alpha_{i} qx) 
 \Phi_{\parallel}(\alpha_{i} )  \\
+ \left[ q_{\nu} ( \delta_{\alpha \mu} - \frac{x_{\mu} q_{\alpha}}{qx}
) - q_{\mu} ( \delta_{\alpha \nu} - \frac{x_{\nu}q_{\alpha}}{qx} ) 
\right]  \int D (\alpha{i}qx)  
 \Phi_{\perp}(\alpha_{i} )     \nonumber
\eea
 Here 
\beq
\int D(\alpha_{i}qx) = \int d \alpha_{1} d\alpha_{2} d \alpha_{3} \delta(
1-\alpha_{1} -\alpha_{2} -\alpha_{3}) e^{-iqx(\alpha_{1} -\alpha_{2} + v 
\alpha_{3} )}
 \eeq
To the leading conformal spin accuracy (i.e. in the asymptotic form) the DA's 
$  \Phi_{\parallel}(\alpha_{i} ) ,  \Phi_{\perp}(\alpha_{i} ) $ are \cite{BF}
\bea
\label{16}
\Phi_{\perp}^{as}(\alpha_{i}) &=& 10 f_{\pi} \delta^2 \alpha_{3}^2 (\alpha_{1} - \alpha_{2} )
                   \nonumber \\
\Phi_{\parallel}^{as} (\alpha_{i}) &=& 0 \eea
where $ \delta^2 $ is defined through the matrix element
\beq
\label{delta}
\lo \bar{d} \gmnu i g \tilde{G}_{\nu \mu} u | \pi(q) \ra = f_{\pi} \delta^2 q_{\mu} \; ,\; \delta^2 \simeq 0.2 \; GeV^2 
\eeq
This matrix element was calculated independently and for different purposes in 
\cite{CZ} and \cite{Nov}.
On the other hand, one can introduce the two-particle DA's of twist 4 by the relation
\bea 
\label{21}
\lo \bar{d}(0) \gmmu \gmf u(x) \rpi = i f_{\pi} q_{\mu} \int_{0}^{1} du e^{-iuqx}
( \phi(u) + x^2 g_{1}(u) ) \nonumber \\
+ f_{\pi}  \left( x_{\mu} - x^2 \frac{ q_{mu}}{qx}
\right) \int_{0}^{1} e^{-iuqx} g_{2}(u)
\eea
where $ \phi(u) $ is the leading twist DA ( with $ \phi^{as} (u) = 6 u \bar{u} $ ) and all logs of $ x^2 $ are included 
in DA's. 
The twist 4 DA's $ g_{1}(u), g_{2}(u) $ can be now expressed in terms of the 
three-particle DA's $ \Phi_{\parallel} , \Phi_{\perp} $. The easiest way to do this is to use the equations of motion \cite{BF}
\bea 
\frac{\partial}{\partial x_{\mu}} \bar{d}(-x) \gmmu \gmf u(x) &=& i \int_{-1}^{1}
dv v \bar{d}(-x) x_{\alpha} g G_{\alpha \beta} (vx) \gamma_{\beta} \gmf u(x)
\nonumber \\
\partial_{\mu} \left( \bar{d}(-x) \gmmu \gmf u(x) \right) &=& i \int_{-1}^{1}
dv  \bar{d}(-x) x_{\alpha} g G_{\alpha \beta} (vx) \gamma_{\beta} \gmf u(x)
\eea
(here $ \partial_{\mu} $ stands for the translation operator). Taking the matrix elements and making use of the definitions (\ref{13}), (\ref{21}) we obtain
to this accuracy
\bea
g_{1}(u) &=& \frac{5}{2} \delta^2 u^2 \bar{u}^2 \nonumber \\
g_{2}(u) &=& \frac{10}{3} \delta^2  u \bar{u} ( u - \bar{u} )  \\
G_{2}(u) &=& \frac{5}{3} \delta^2 u^2 \bar{u}^2  \nonumber
\eea
For further convenience we have introduced here the DA $ G_{2}(u) $ defined by $ g_{2} (u) = (d/du)G_{2}(u) $. Now we are in position to show that for $ b^2 \equiv - x^2 $ the combination $ b^2 u \bar{u} $ is exactly the expansion parameter in the coordinate space. To see this, we choose the frame with $ x= 
(x_{+},x_{-},x_{\perp}) = (0,x_{-},x_{\perp}) $. Then
\bea
\label{24}
\lo \bar{d}(0) \gamma_{+} \gmf u(0,x_{-},x_{\perp}) \rpi =  i f_{\pi}q_{+}
\int_{0}^{1} du e^{-iuq_{+}x_{-}} ( \phi(u) + x^2 ( g_{1} + G_{2})(u) \nonumber
\\
=    i f_{\pi}q_{+}
\int_{0}^{1} du e^{-iuq_{+}x_{-}} 6u \bar{u} ( 1 - \frac{5}{9} \delta^2   b^2 u \bar{u} + \cdots )
\eea
It is seen that we have obtained the two first terms in the expansion of the pion WF $ \Psi(\xi=
2 u -1, b^2)  $ in powers of $ b^2 $. It is now easy to find the value of $ \la \k \ra_{\pi} $ from Eq.(\ref{24}) :
\beq
\label{25}
\lo \bar{d} \gamma_{+} \gmf ( i \nabright_{\perp} )^2 u \rpi = i f_{\pi} q_{+}  \frac{5}{9}  \delta^2 \equiv  i f_{\pi}  q_{+}    \la \k \ra_{\pi}   \; , \;   
\la \k \ra_{\pi}  \simeq (330 \; MeV)^2 
\eeq
We have thus demonstrated two important facts:  \\
1. The light-cone WF $ \wfu $ and higher twist DA's containing gluons explicitly are two sides of the same coin, as they are related by exact equations of motion.  \\
2. The combination $ \k / u \bar{u} $ is a single variable which the WF depends on :   \\
$ \wfu = \Psi ( \k / u \bar{u} ) $.  \\
The same line of reasoning can be extended to the case of the $\rho$-meson. We therefore conjecture that  the ansatz (\ref{12}) is valid in the whole range of $ \xi $ to the leading conformal 
spin accuracy, i.e. it corresponds to a series of asymptotic two-particle DA's 
of higher twists. This approximation seems reasonable provided the WF $ \wf $
is evaluated at a high normalization point $ \mu^2 \sim $ a few $ GeV^2 $. 
On the other hand, it is such $ \mu^2 $ that enters Eq.(\ref{12}) following
from the duality arguments.

\clearpage
\appendix
\def\theequation{\thesection.\arabic{equation}}

\section*{Appendix B}

\def\thesection{B}
\setcounter{equation}{0}
The aim of this Appendix is to present 
the method \cite{Zh94} which allows to estimate some high-dimensional
condensates with a reasonable accuracy. The importance of this question
(apart from  the pure theoretical   
interest concerning a structure of the 
 QCD vacuum) for the phenomenological purposes was emphasized
in Section 3. It was observed   that 
the large magnitude for the  high dimensional condensates
implies a strong momentum fluctuations in the transverse direction.
Correspondingly, it has a remarkable 
impact on  the pre-asymptotic behavior of the electroproduction.

Let us start with the formulation of the idea exploited in this method. Consider some 
correlation function $i\int dx e^{iqx}
\la T\{J_1(x),~J_2(0)\}\ra$ at large $-q^2$.
If the currents $J_1, J_2$ are chosen in such a way
that in the chiral limit the perturbative contribution is zero,
we end up with the leading contribution in the form
$\frac{\la O\ra}{q^2}$ plus some nontrivial function, but with small
numerical coefficient
due to the loop suppression. Such a behavior, from the point 
of view of the dispersion relations (where each resonance $|r\ra$
contributes  to the correlation function 
as $\frac{1}{-q^2+m_r^2}\cdot \la 0|J_1|r\ra \la r|J_2|0\ra$)
 implies that in the chiral limit we can
keep only the $\pi$ meson contribution and, therefore,
we make an identification
$\la O\ra \simeq \la 0|J_1|\pi\ra \la \pi|J_2|0\ra$. An accuracy
of this equation can be estimated by calculating  corrections
to the leading term which is $\sim\frac{\la O\ra}{q^2}$.
These corrections,  which usually do  not exceed
$10\div 20\%$, are responsible for the 
contribution of the higher states 
and proportional to  $\sim \la 0|J_1|r\ra \la r|J_2|0\ra$.
  Those contributions
have the same order of magnitude as the corrections in OPE, 
  i.e. about $10\div 20\%$. If it were not true,
  the  matching of the right hand side and
left hand side of the corresponding sum rules would  be not   possible. 
Therefore, if we knew the $\pi$-meson matrix elements 
$\la 0|J_1|\pi\ra \la \pi|J_2|0\ra$ independently (from PCAC, for example),
 we would  estimate the condensate $\la O\ra$ with a high enough accuracy. 

We demonstrate the method by considering the following correlation
function
\beq
\label{1a}
i\int dx e^{iqx}
\la T\{J_1^{\mu}(x),~J_2(0)\}\ra =q^{\mu}T(q^2),~~\nonumber \\
J_1^{\mu}=\bar{d}\gamma^{\mu}\gamma_5 g\sigma Gu ,~~
J_2=\bar{u}\gamma_5 g\sigma Gd
\eeq
where $\sigma G\equiv \sigma_{\mu\nu}G^a_{\mu\nu}\frac{\lambda^a}{2}$ and $ \sigma_{\mu\nu} = i/2 ( \gmmu \gmnu - \gmnu \gmmu) $ .
The leading contribution is determined by the condensate
of interest:
\beq
\label{2a}
T(q^2)=\frac{1}{q^2}\la \bar{q}(g\sigma G)^2q\ra .
\eeq

At the same time, both $\pi$ meson matrix elements which enter this correlator are known-- one of them 
$\la 0|\bar{d}\gamma^{\mu}\gamma_5g\sigma Gu|\pi\ra$
can be expressed in terms of the matrix element (\ref{delta}) which we already know, 
and another one $\la \pi|\bar{u}\gamma_5 g\sigma Gd|0\ra$
can be found from PCAC.
Finally we arrive at the following
equation:
\beq
\label{3b}
\la \bar{q}g^2(\sigma G)^2q\ra = 4\delta^2\cdot\la\bar{q}(g\sigma G)q\ra
\simeq \frac{\la\bar{q}(g\sigma G)q\ra^2}{\la
\bar{q}q\ra}\simeq m_0^4 \la\bar{q}q\ra , 
\eeq
where we used the following $\pi$ meson matrix element
\beq
\label{3c}
\la 0|\bar{d}g\tilde{G}_{\mu\nu}\gamma_{\nu}u|\pi(q)\ra
=if_{\pi}q_{\mu} \delta^2\simeq
if_{\pi}q_{\mu}\frac{\la\bar{q}(g\sigma G)q\ra}{4\la
\bar{q}q\ra},~~ \delta^2\simeq 0.2 \; GeV^2.
\eeq
This matrix element was calculated 
in Ref.\cite{Nov} (in terms of $\delta^2$ )and in Ref.\cite{CZ} 
(in terms of mixed vacuum condensate). Numerically these results are
in the perfect agreement to each other.

As we mentioned, an accuracy of Eq.(\ref{3b}) can be estimated by 
 calculating   
  corrections to the leading term (\ref{2a}). 
The most important loop correction can be calculated explicitly \cite{Zh94} and
it turns out to be  30 times smaller than the main contribution (\ref{2a}); therefore,
the precision of our estimate (\ref{3b}) for the condensate 
$\la \bar{q}g^2(\sigma G)^2q\ra$ is determined mainly by the 
accuracy of the $\pi$ meson matrix elements which, we believe,
is reasonably  high.
If we describe a deviation from the factorization prescription for the condensate
$\la \bar{q}g^2(\sigma G)^2q\ra$ by introducing the parameter
$K$ as a measure of the non-factorizability ($K=1$ 
if the factorization would work), we  get the following estimate from Eq.(\ref{3b})
for the parameter $K$:
\beq
\label{4a}
\la \bar{q}g^2(\sigma G)^2q\ra = \frac{K}{3}
\la g^2G_{\mu\nu}^2\ra \la\bar{q}q\ra,~~K\simeq 3\div 4.
\eeq  
which is the main result of these calculations.
To be on the safe side we use $K\simeq 3$ in all our numerical
estimates in the text.

\end{document}